# Turbulence Momentum Transport and Prediction of the Reynolds Stress in Canonical Flows

T.-W. Lee

*Mechanical and Aerospace Engineering, Arizona State University, Tempe, AZ, 85287*

**Abstract-  We present a unique method for solving for the Reynolds stress in turbulent canonical flows, based on the momentum balance for a control volume moving at the local mean velocity.  A differential transform converts this momentum balance to a solvable form. Comparisons with experimental and computational data in simple geometries show quite good agreements.  An alternate picture for the turbulence momentum transport is offered, as verified with data, where the turbulence momentum is transported by the mean velocity while being dissipated by viscosity.  The net momentum transport is the Reynolds stress. This turbulence momentum balance is verified using DNS and experimental data.**

T.-W. Lee
Mechanical and Aerospace Engineering, SEMTE
Arizona State University
Tempe, AZ 85287-6106
Email: attwl@asu.edu

**Nomenclature**

$C_1$        =  constants
$Re_\tau$        =  Reynolds number based on friction velocity
$U$        =  mean velocity in the x direction
$U_e$        =  free-stream velocity
$V$        =  mean velocity in the y direction
$u'$        =  fluctuation velocity in the x direction
$u_{rms}'$        =  root-mean square of u'
$u'v'$        =  Reynolds stress
$\delta$        =  boundary layer thickness
$\nu_m$        =  modified kinematic viscosity

## Introduction

Finding the Reynolds stress has profound implications in fluid physics. Many practical flows are turbulent, and require some method of analysis or computations so that the flow process can be understood, predicted and controlled. This necessity led to several generations of turbulence models including a genre that models the Reynolds stress components themselves, the Reynolds stress models [1, 2]. The conventional approach of writing higher moments for the Reynolds stress, then attempting to model them, involves ever increasing number of terms and corresponding complexities. For example, in the Reynolds stress budget alone, there are seven (five) groups of (source/sink) terms [3], which when expanded become about three times the number. In addition, one of the five groups of terms is the dissipation rate, which in itself has quite a complex budget [3]. Thus, it is of some interest to be able to find the Reynolds stress in terms of root turbulence variables, in a simple, physics-based approach. From a fundamental perspective, prediction of the Reynolds stress from the first principles would also be quite significant. The current work is unique in that it is a new theoretical development, and not related to any of the previous works. Also, we make no claim to review the vast literature in turbulence models, so we refer to the works cited in Ref. 4 for those interested in turbulence modeling. As for theoretical work, some of the turbulence modeling work [5] including the Reynolds stress models [1, 2] start out as analytical or theoretical. However, soon due to the complexity and sheer number of unknown terms, modeling of the terms becomes unavoidable [1-5]. Early theoretical works tend to be quite complex [6-10], and/or their applicability to general turbulent flows is doubtful if not impossible.

Recently, we developed a theoretical basis for determination of the Reynolds stress in canonical flows [11-13]. It is based on the turbulence momentum balance for a control volume moving at the local mean flow speed. The resulting "integral formula" works quite well in determining the Reynolds stress based on inputs of root turbulence parameters, such as streamwise component of the turbulence kinetic energy, the mean velocity and its gradient. The predicted Reynolds stress is in good agreement with experimental and DNS data [11-13]. This approach opens the possibility for broad applications in turbulence analyses and simulations. Here, we provide further validations and explanations of the turbulence momentum balance, so that groundwork for applications in more complex, three-dimensional flows is founded.

**Turbulence Momentum Transport**

The derivation of the turbulence momentum balance has been presented elsewhere [11-13], but to place this work in context we briefly reiterate the main steps and also refer to Appendix. For a control volume moving at the local mean speed, as shown in Figure 1, the momentum balance can easily be written as:

$$\frac{\partial (u'v')}{\partial y} = -\frac{\partial u'^2}{\partial x} + \nu_m \frac{\partial^2 u_{rms}'}{\partial y^2} \tag{1}$$

The terms in Eq. 1 are analogous to the mean momentum equation, and reflect the fact that in a coordinate frame moving at the local mean velocity only the fluctuation terms will be observed. The gradient of the pressure fluctuation is expected to be significant only for compressible flows, so we omit the pressure fluctuation term for "incompressible" turbulence at low Mach numbers [11-13]. As will be shown, these assumptions lead to viable results. In conventional calculations in the absolute coordinate frame, the x-derivatives would have been set to zero for fully-developed flows, and we would be left with a triviality. However, for a boundary-layer flow as an example (Figure 2), the boundary layer grows due to the "displacement" effect. The mass is displaced due to the fluid slowing down at the wall, as is the momentum, and turbulence parameters as well. The boundary layer thickness grows at a monotonic rate, depending on the Reynolds number. Thus, if one rides with the fluid moving at the mean velocity, one would see a change in all of the turbulence properties, as illustrated in Figure 2. This displacement effect can be mathematically expressed as:

$$\frac{\partial}{\partial x} = C_1 U \frac{\partial}{\partial y} \qquad (2)$$

I.e., the fluid parcel will see a different portion of the boundary layer in the y-direction, and how much difference it will see depends on how fast the fluid is moving along in the boundary layer. Thus, the mean velocity, U, appears as a multiplicative factor in Eq. 2. $C_1$ is a constant that depends on the Reynolds number.

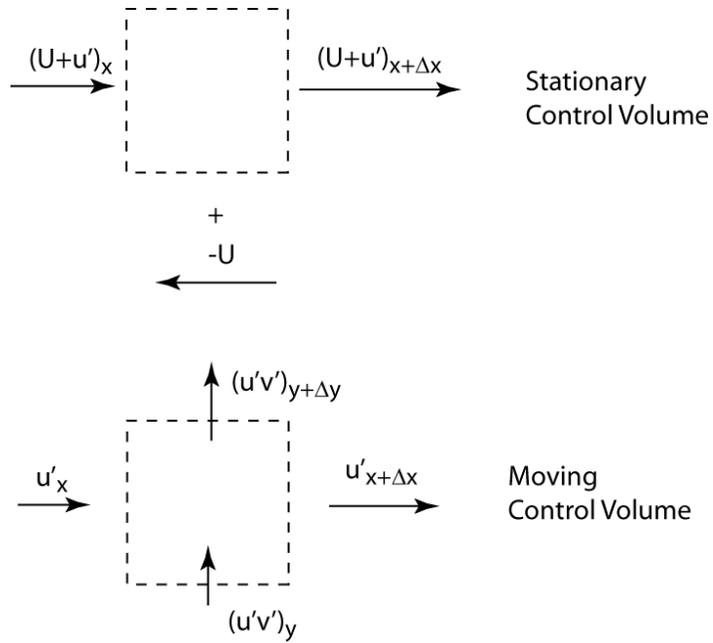

**Figure 1. Illustration of the momentum balance for a moving control volume.**

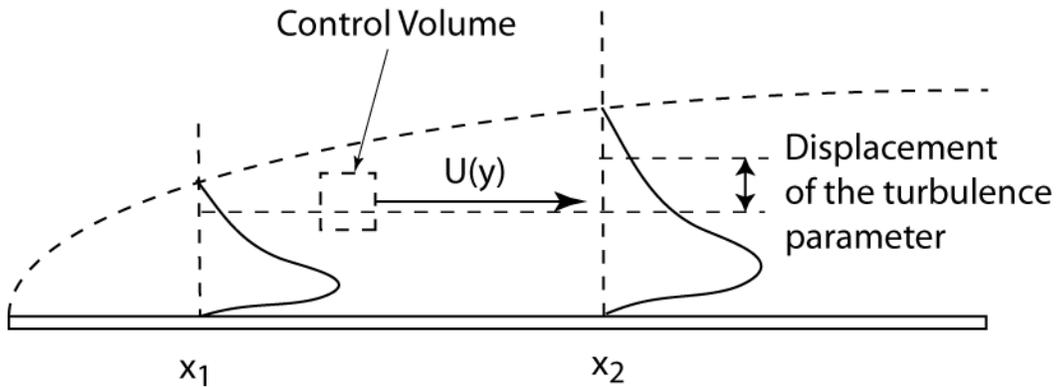

**Figure 2. A schematic illustration of the concept of the displaced control volume method.**

Eqs. 1 and 2 lead to the "turbulence momentum equation".

$$\frac{\partial (u'v')}{\partial y} = -C_1 U \frac{\partial u'^2}{\partial y} + \nu_m \frac{\partial^2 u_{rms}'}{\partial y^2} \tag{3}$$

$v_m$ is the modified viscosity, the value of which is close to but not equal to the laminar viscosity. This will later be discussed in the context of some results. Eq. 3 would indicate, as will be proven, that complex modeling of the Reynolds stress is not necessary, and that we can use the turbulence momentum balance to determine the Reynolds stress from root turbulence parameters [11-13]. Only U and $u'^2$ are needed to find the gradient of u'v'. This is a far simplification than the conventional budget for the Reynolds stress which includes seven groups of terms (e.g. production, diffusion, dissipation, etc), that lead to many more when expanded in x- and y-directions. Eq. 3 can be numerically integrated based on appropriate boundary condition, to find the Reynolds stress. For complex flows, the full array of $u_iv_j$ terms can be determined in this manner, which is currently being investigated [14]. Eq. 3 is a relatively simple ordinary differential equation, which can be integrated. We have referred to the integrated form as the "integral formula", and it gives explicitly the Reynolds stress that compares well with DNS and experimental data [11-13]. But this integration introduces an integration constant and also a non-local term (an integral term). Here, we examine the turbulence momentum balance as expressed in Eq. 3, and the implications of the terms therein.

**RESULTS AND DISCUSSION**

We can verify the validity of the turbulent momentum balance in Eq. 3. We first obtain the left-hand side (LHS) of Eq. 4 from u'v' in the DNS data for rectangular channel flows [15, 16], by computing its derivative $\frac{\partial (u'v')}{\partial y}$ numerically. In Iwamoto et al. [15, 16], DNS data for $Re_\tau$ up to 590 are presented. We can use this data to compute the terms on the RHS of Eq. 3. Only the data for U and $u'^2$ are needed, since $u'_{rms}$ is equal to the square root of $u'^2$ in this formulation. The derivatives are computed using a second-order method. This comparison is shown in Figure 3, where the LHS and RHS of Eq. 3 are quite well balanced, in spite of the fact that we are taking first- and second-derivatives of discrete data [16]. Therefore, the conservation of turbulence momentum (Eq. 3) holds for channel flows. This has some implications for analysis and computations of turbulent flows: complex models for the Reynolds stress are not necessary, and there is a simple momentum balance that leads to an explicit expression for the Reynolds stress (gradient) (Eq. 3).

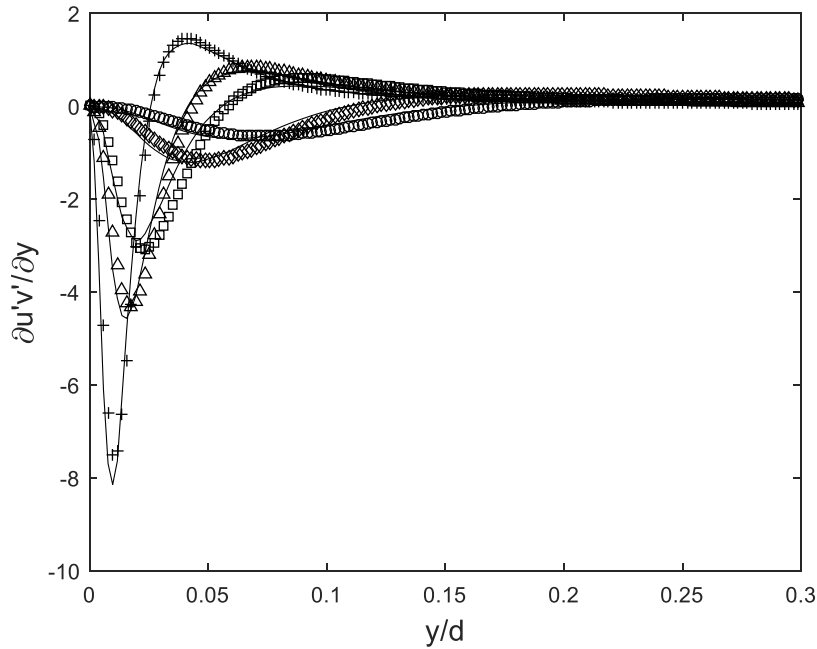

**Figure 3. Comparison of the gradient of the Reynolds stress for rectangular channel flows.**

**Lines represent current results (Eq. 3).**

**Data symbols: circle (Re$_\tau$ = 110), diamond (150), square (300), triangle (400), + (650) [16].**

The dynamics of the turbulence momentum transport can be examined through the contributing terms to the Reynolds stress, i.e. the RHS of Eq. 3. The mean velocity in channel flows exhibits the familiar, steep profile near the wall, and then tending to the centerline value as one moves away from the wall. We plot u'$^2$ from Iwamoto et al. [15, 16], in Figure 4 for visualizing the transport process. We can see that $u'^2$ profiles have a peak near the wall, then gradually approach the centerline value. Finding the right combination of U, $u'^2$, and numerous other parameters for the Reynolds stress through artificial means spawned an entire library of turbulence models. However, we can see (in a hindsight) that there are some hints such as the approximate similarity in $u'^2$ profiles with the Reynolds stress, except for the inverted sign (+/-), and that $u'^2$ profiles exhibit a sharp peak near the wall. In Figure 5, we plot the two terms from the RHS of Eq. 3, which we refer to as the transport and viscous term, respectively. The gradient of the Reynolds stress is also plotted (solid line) for $Re_\tau = 650$. The transport term contains the product of U and $u'^2$ gradient. The viscous term is important only near the wall, and it is this term that adds to the transport term to give a sharp negative gradient of the Reynolds stress in Figure 5. Convexity upward in u' (taken as the square root of $u'^2$) leads to negative second derivative, and therefore the viscous term adds to the negative gradient in the Reynolds stress in Eq. 3. Away from the wall, the transport term, $-C_1 U \dfrac{\partial u'^2}{\partial y}$, overlaps with the Reynolds stress since the viscous term is significant only near the wall. That is, the slope of $u'^2$ is nearly constant in Figure 4 away from the wall, causing the second gradient to become nearly zero. Figures 4 and 5 demonstrate that Eq. 3, in its simple form, contains the correct physics of turbulence momentum transport. The Reynolds stress is simply turbulence transport of momentum, and obey the same conservation principle.

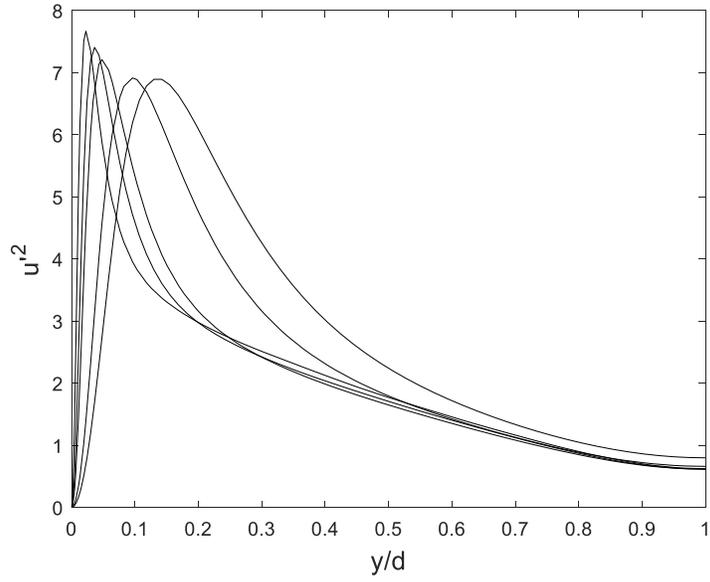

**Figure 4.  u'² profiles from Iwamoto et al. [16] for Re_τ = 110 - 650.**

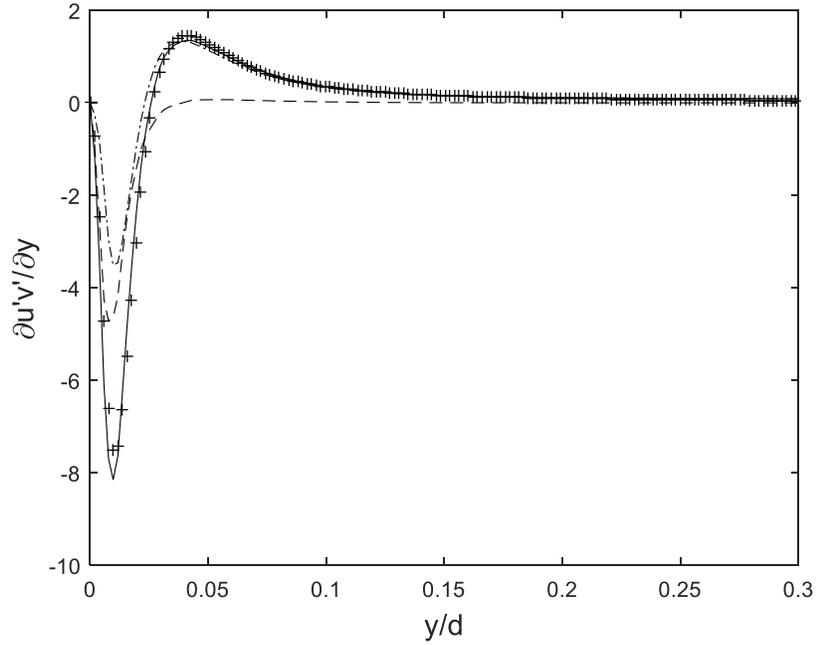

**Fig. 5.  Contributing terms to the Reynolds stress gradient, $\frac{\partial(u'v')}{\partial y}$, in Eq. 4.  Re_τ = 650.**

**Data symbol:  $\frac{\partial(u'v')}{\partial y}$ ; dash-dot:  $-C_1 U \frac{\partial u'^2}{\partial y}$ ; and dashed:  $\nu_m \frac{\partial^2 u_{rms}'}{\partial y^2}$.**

The constant, $C_1$, in Eq. 3, arises due to the displacement effect as described in Eq. 2. We expect this displacement effect to be larger for higher Reynolds numbers, so that we also expect $C_1$ to increase with increasing Reynolds number. $C_1$ ranged from 0.000536 at Ret = 110 to 0.0012 at Ret = 650. The modified viscosity, $\nu_m$, is close but not equal to the laminar viscosity. The reason is that this viscous term arises due to mainly due to fluid shear, and only partially molecular transport depending on the Reynolds number. Thus, a modification to the viscosity is needed, and thus the use of the term "modified viscosity" for $\nu_m$. We expect that this viscous effect occurs over a range of scales in turbulent flows, and the average length and velocity scales such as the Taylor microscale will be related to the modified viscosity.

We can perform a similar verification for planar jets, for which we use the experimental data of Gutmark and Wygnanski [17, 18]. For free jets, Eq. 3 becomes yet simpler because there are no solid boundaries and sharp gradients are not present in $u'^2$. This causes the viscous term in Eq. 3 to be negligible except near the peak of $u'^2$ profiles, and the Reynolds stress is mostly determined by the transport term, $-C_1 U \dfrac{\partial u'^2}{\partial y}$. Figure 6 shows the gradient of the Reynolds stress from Gutmark and Wygnanski [17, 18], compared with current "theory" using U and $u'^2$ data. The initial and final slope of the Reynolds stress is not quite matched, but overall the agreement is good. $u'^2$ increases (positive slope) from the centerline value, and then decreases (negative slope), while the mean velocity gradually decreases from the centerline value to zero [17, 18]. With $C_1$ being negative (to follow the sign convention for u'v' in jets), this leads to a positive gradient in the Reynolds stress near the centerline, then converting to negative gradient.

As noted above, the RHS side of Eq. 3 can be integrated to find the Reynolds stress itself, starting from the boundary of condition of u'v' = 0 at the centerline. The result is shown in Figure 7, where we can see that the integration is somewhat forgiving of the discrepancy in the initial slope. The agreement for the Reynolds stress is decent in Figure 7, although the error does tend to accumulate particularly beyond the negative peak near y/y$_{m/2}$ = 0.75. Again, we are taking the derivative of experimental data in Eq. 3, fitted to a line [18], and considering the potential for numerical errors, the comparisons of Figures 6 and 7 are quite reasonable. We note that when plotted in self-similar variables, both the data and theoretical line collapse to a single profile [17, 18].

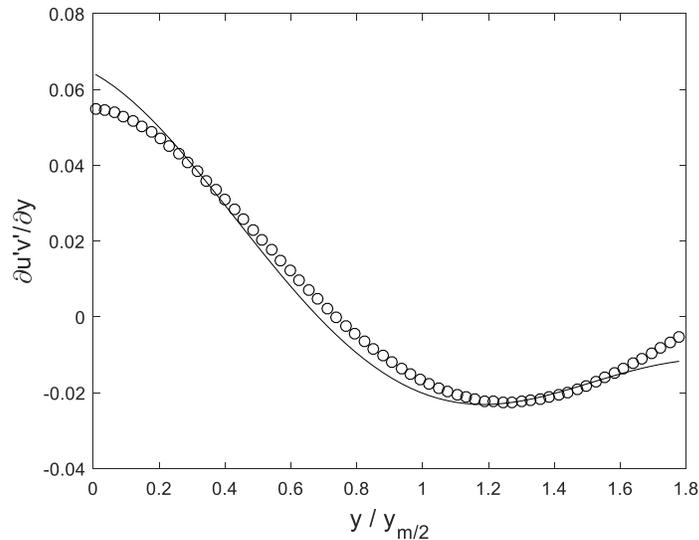

**Figure 6. Comparison of the gradient of the Reynolds stress for planar jets. The data (symbol) are from Gutmark and Wygnanski [17, 18]. The line represents current result (RHS of Eq. 3).**

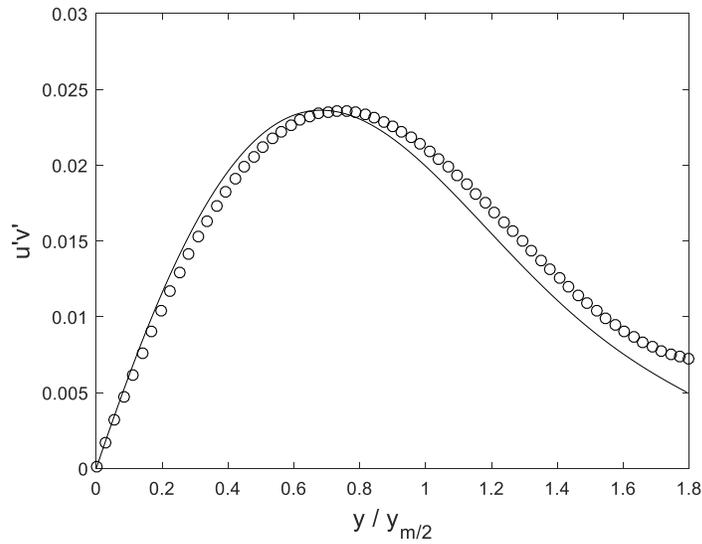

**Figure 7. Comparison of the Reynolds stress for planar jets.**

**The data (symbol) are from Gutmark and Wygnanski [17, 18]. The line represents**

**current result by numerically integrating the RHS of Eq. 3.**

Finally, we use the DNS data for flows over a flat plate at zero pressure gradients [19]. $u'^2$ profiles have quite a sharp peak, near the wall, and gradually decrease as the distance from the wall increases [19]. Although this is somewhat attenuated near the wall by the mean velocity in the transport term (the first term on the RHS of Eq. 3), the resulting contribution of $u'^2$ is still high. However, close to the wall the large gradient in $u'^2$ also increases the magnitude of the viscous term. For experimental data [19], there are fluctuations due to the short distance over which turbulence parameters change, and this leads to large errors when numerically differentiated. So we apply smoothing spline-fit functions, to remove these fluctuations. Otherwise, there are some "leakages" in the numerical analysis [11-13]. Once the spline fit is obtained and used, the balance in the turbulence momentum can be checked by again integrating the RHS to find the Reynolds stress directly and compare with experimental data [19]. The results are shown in Figure 8. In spite of the sharp gradients in $u'^2$ near the wall, the agreement between the current theory and experimental data is quite good. Similar to the channel flow, the transport term is pervasive while the viscous term adds to the Reynolds stress near the wall. The viscous term is negative ($u'_{rms}$ convex upward) near the wall, which adds to the transport term. When integrated to find the Reynolds stress from Eq. 3, the viscous term adds to the transport term, which follows the general shape of the Reynolds stress. This is so-called cumulative or the displacement effect, also observed through the integral formula in our first work of this kind [11]. Overall, observations in channel and flat plate flows indicate that the transport for the Reynolds stress is similar in wall-bounded flows.

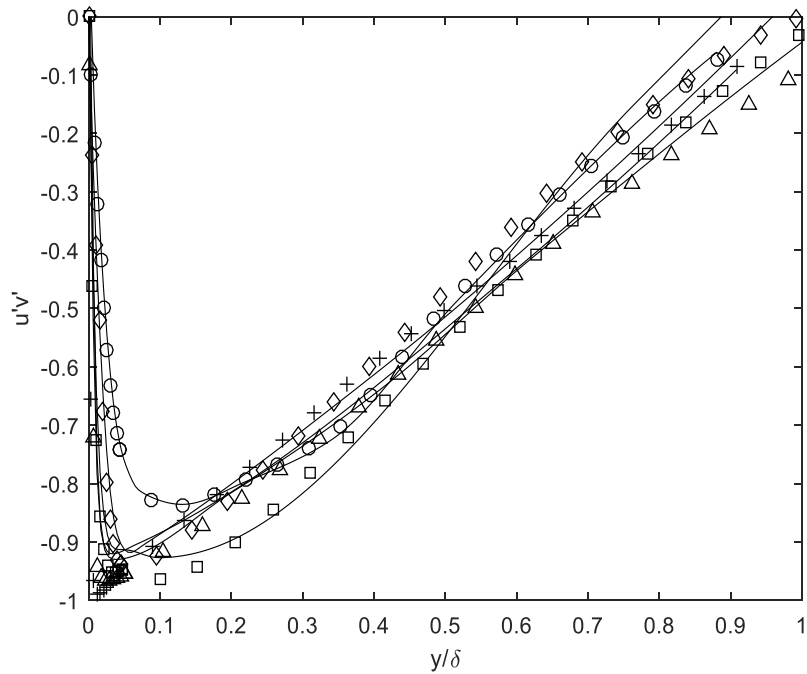

**Figure 8. Comparison of the Reynolds stress in boundary layer flows over a flat plate. The data (symbols) are the Reynolds stress for Re$_\theta$ = 1430 - 31000 [19]. Lines represent RHS of Eq. 3 integrated.**

**CONCLUSIONS**

The current approach of applying the momentum balance to a control volume moving at the local mean velocity is a novel and a simple approach, which yields a closed-form equation for the Reynolds stress gradient. This formulation is confirmed using various data sets, where the gradient of the Reynolds stress is shown to be due to the net momentum balance of the transport and the viscous effects. This gradient can be numerically integrated to find the Reynolds stress itself. Both the gradient and the Reynolds stress obtained through this approach agree quite well with DNS and experimental data in canonical flows. Implications of this work are that the Reynolds stress can be written explicitly in terms of basic turbulence parameters in a simple form, derived from pure fluid physics, and that potential exists for applications of this concept to more complex turbulent flows.

# Appendix: Explanation of the Turbulence Momentum Equation

In Reynolds-averaged Navier-Stokes (RANS) equation, non-linear terms involving turbulent fluctuation velocities arise since the absolute velocity is decomposed into mean (U) and fluctuating (u') component: $u = U + u'$. The non-linear terms that develop during the averaging process in the RANS are called the Reynolds stress, which involves time-averaged components of products of fluctuating velocities, $u_i' u_j'$. Here, we omit the bar above u'², u'v', etc, for simplicity, and take all the fluctuation parameters to be time-averaged. Figuring out how the Reynolds stress is related to the mean and other "root" turbulent parameters has been the topic of numerous studies, for quite some time. However, we notice that the decomposition is necessary only in the absolute (stationary) coordinate frame, as shown in the top part of Figure 1. What if we consider a control volume moving at the mean flow? Then, an observer riding this control volume will not "feel" the mean flow, but only the fluctuating components, as shown in the Figure 1. We can re-do our momentum balance for this control volume, where only the root-mean-square (rms) fluctuations of the velocities (momentum) need to be balanced, in the x-direction. Momentum balance in other directions would produce other components of the Reynolds stress. The momentum equation is greatly simplified in this relative coordinate frame:

$$\frac{\partial u'^2}{\partial x} + \frac{\partial (u'v')}{\partial y} = \nu \frac{\partial^2 u_{rms}'}{\partial y^2} \qquad (A1)$$

For stationary flows, the fluctuating momentum gradients are balanced by the viscous force term, and the pressure term is neglected. By using the Lagrangian control volume, the mean momentum terms are de-coupled from the fluctuating terms. We also apply the boundary layer approximation (V<<U) and neglect the laminar viscous term based on gradient of U.  Again, all the quantities ($u_{rms}$', $u'^2$, $u'v'$) in Eq. A1 are time-averaged.  The idea of time averaging in this framework is relatively straight-forward for all of the terms, except for the viscous term.  The viscous shear stress is caused by instantaneous gradient of the u' in the y-direction.  Thus, the concept here is that there can be some time-averaged gradient (and second gradient) in u' that would then cause the viscous shear stress.

After applying the differential transform (Eq. 2) to account for the displacement effect, we obtain Eq. 3.

**ADDENDUM**

From the submission of the earlier work, some further developments are added.

We can write an equation for transport of $u'^2$.

$$C_2 U \frac{\partial (u'^3)}{\partial y} = -\frac{\partial (u'^2 v')}{\partial y} + \nu_m \left( \frac{\partial u_{rms}'}{\partial y} \right)^2 \tag{4}$$

In this formulation, the source and sink terms are simplified to just the viscous dissipation term in Eq. 3.  The effectiveness of this approach can be judged by the results below.  Eqs. 3 and 4 then are the transport equations for turbulence momentum and energy (in the x-direction), and represents a closable set to find the Reynolds stress explicitly.  Eqs. 3 and 4 are direct results of applying the conservation principles for a control volume moving at the local mean velocity.

We verify the validity of the turbulent transport equations in Eq. 4, as shown in Figure 4, where the LHS of Eq. 4, $C_2 U \dfrac{\partial (u'^3)}{\partial y}$ (plotted with data symbols) is compared with the RHS (lines) for various Reynolds numbers.  Here, we equate the dreaded higher-order term u'$^2$v' with u'v'*u', or the Reynolds stress times u'$_{rms}$.  This is an approximation that serves to suppress the introduction of another unknown variable v'$_{rms}$.  However, comparisons with data (Figures 4 and 5) show that this approximation works reasonably well.  After all, Eq. 4 only attempts to capture the transport of u'$^2$, and this is accomplished by u'$^2$v' in the y-direction.  The initial peak is closely matched by the RHS of the transport equation at all Reynolds numbers, while the second negative peak is underestimated at higher Reynolds numbers.  The locations of these peaks are nearly perfectly picked off by Eq. 4.  The underestimation of the second peaks may be due to simplification of the source terms into a single dissipation term, or that the turbulent kinetic energy in the x-direction (u'$^2$) cannot be individually balanced as in the momentum transport.  The viscous dissipation cannot distinguish the kinetic energy components in x- or y-direction.  For the time being, we retain these approximations to demonstrate that a set of equations to solve for Reynolds stress can be formulated, and refinements to Eq. 4 will improve the accuracy of the transport equation.

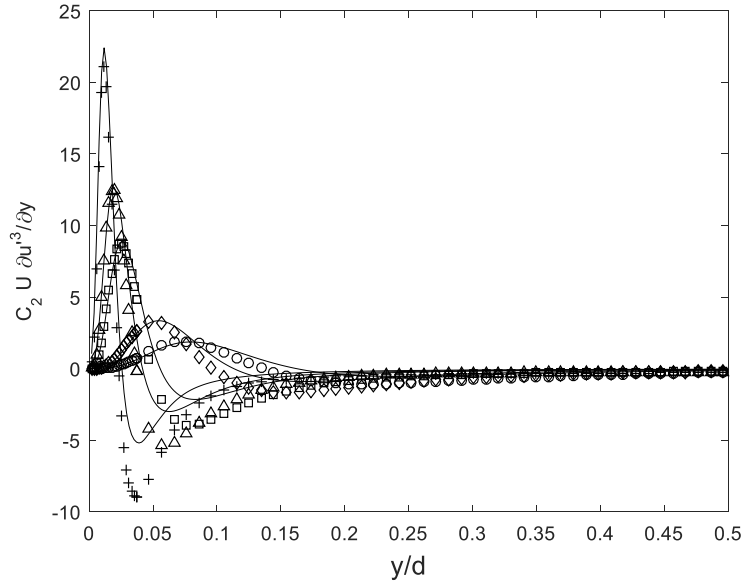

**Figure 4. Validation of the u'² transport (Eq. 4) for channel flows.**

**Data symbol:** $-C_2 U \dfrac{\partial u'^3}{\partial y}$; **lines: RHS of Eq. 4. Same data symbols as in Figure 2 are used.**

We can perform a similar verification for planar jets, for which we use the experimental data of Gutmark and Wygnanski [7, 8]. For free jets, Eq. 3 becomes yet simpler because there are no solid boundaries and sharp gradients are not present in u'². This causes the viscous term in Eq. 3 to be negligible except near the peak of u'² profiles, and the Reynolds stress is mostly determined by the transport term, $-C_1 U \dfrac{\partial u'^2}{\partial y}$. Figure 5 shows the gradient of the Reynolds stress, and also comparison of the terms in Eq. 4, LHS = $-C_2 U \dfrac{\partial u'^3}{\partial y}$ vs. RHS of Eq. 4. The initial and final slope of the Reynolds stress is not quite matched, but overall the agreement is good. u'² increases (positive slope) from the centerline value, and then decreases (negative slope), while the mean velocity gradually decreases from the centerline value to zero [7, 8]. With $C_1$ being negative, this leads to positive gradient in the Reynolds stress near the centerline, then converting to negative gradient. Comparison of the $u'^2$ transport terms in Eq. 4 is also not perfect, but given the fact that we are taking up to second-derivatives of experimental data, the agreements are reasonable and the overall dynamics of the turbulent transport appear to be captured.

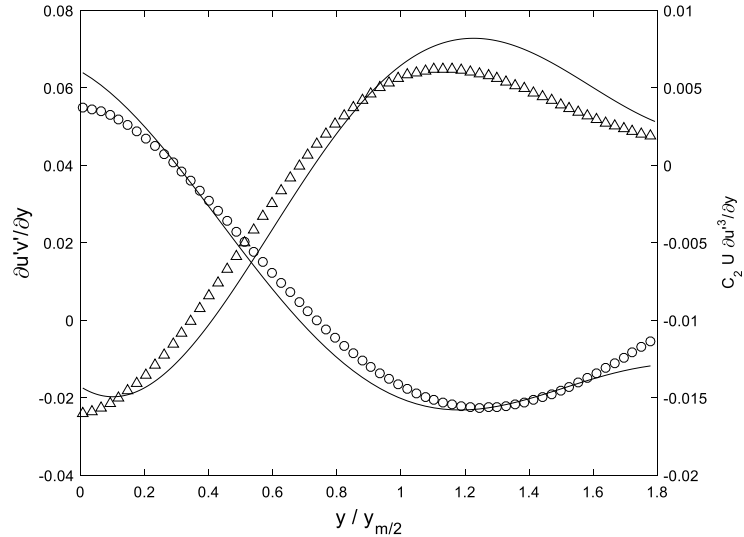

**Figure 5. Validation of the transport equations, Eq. 3 (LHS = circles) and Eq. 4 (LHS = triangles) for rectangular jets. Lines are the RHS's. The data from Gutmark and Wygnanski [7, 8] have been used.**

**SUMMARY**

We have found a simple format to find the Reynolds stress in canonical flows, based on the turbulence momentum and kinetic energy ($u'^2$) balance for a control volume moving at the local mean flow velocity. This set of transport equations is verified using experimental and DNS data, with good agreements and logic. It furnishes two equations, to solve for two unknowns, u'v' and $u'^2$, and in conjunction with the Reynolds-averaged Navier-Stokes equations, turbulent flow field can be determined, at least in canonical geometries at this point. There is room for improvements in the $u'^2$ equation (Eq. 4), but this work is the first in presenting a solvable set of equations for turbulence.